\documentstyle[12pt,aaspp4]{article}

\begin{document}
\title{Limits on Hot Intracluster Gas Contributions to the 
Tenerife Temperature Anisotropy Map}

\author{J.A. Rubi\~no-Mart{\'\i}n}
\affil{Instituto de Astrof{\'\i}sica de Canarias\\
	V{\'\i}a L\'actea s/n. 38200 La Laguna, Spain.\\
email: jalberto@ll.iac.es}

\author{F. Atrio-Barandela, C. Hern\'andez-Monteagudo} 
\affil{F{\'\i}sica Te\'orica. Facultad de Ciencias.\\
        Universidad de Salamanca, 37008 Spain.\\
email: atrio@astro.usal.es, chm@orion.usal.es}

\begin{abstract}
We limit the contribution of the hot intracluster gas, 
by means of the Sunyaev-Zel'dovich effect, to the
temperature anisotropies measured by the Tenerife experiment.
The data is cross-correlated with 
maps generated from the ACO cluster catalogue, the ROSAT PSPC
catalogue of clusters of galaxies, a catalogue of 
superclusters and the HEAO 1 A-1 
map of X-ray sources. There is no evidence of contamination
by such sources at an rms level of $\sim 8\mu$K at 99\% 
confidence level at $5^o$ angular resolution. We place an upper limit 
on the mean Comptonization parameter of $ y \le 1.5\times 10^{-6}$
at the same level of confidence. 
These limits are slightly more restrictive than those previously
found by a similar analysis on the COBE/DMR data and
indicate that most of the signal measured by Tenerife is cosmological.

\end{abstract}


\section{Introduction}

The discovery of Cosmic Microwave Background (CMB) temperature
anisotropies (Smoot et al. 1992; Wright et al. 1992) immediately 
prompted the question of their origin. In most cosmological scenarios
temperature anisotropies are expected to arise
with the growth of matter density perturbations (Peebles 1980);
but it was also suggested (Hogan 1992) that
the anisotropies could have been originated by
inverse Compton scattering from hot diffuse clouds of
electrons in nearby superclusters (see Birkinshaw 1999
for a review on the Sunyaev-Zel'dovich effect). 
Right after Hogan's suggestion, Boughn \& Jahoda (1993)
and Bennett et al. (1993) searched for non cosmological signal
contributions to the COBE/DMR 2 year sky maps. If the Sunyaev-Zel'dovich (SZ)
component is significant at some level and the distribution of
hot electrons is traced by rich 
clusters or any other extragalactic source, then temperature 
anisotropies should be correlated with maps constructed from
extragalactic source surveys.
Banday et al. (1996) performed this analysis for the
COBE/DMR  4 year map. Their study did not find
a statistically significant contribution, reassuring the idea that
temperature anisotropies were of cosmological origin.
Similar cross-correlation analysis have been carried
out by several authors to estimate the rms level of
the Galactic emission present in different CMB maps
(Kogut et al. 1996; Kneissl et al. 1997; 
de Oliveira-Costa et al. 1997, 1998, 1999).

In this article we  study the contribution of
nearby structures on the temperature anisotropies measured by the
Tenerife map (Guti\'errez et al. 2000). 
Two effects associated with the local matter
distribution can give rise to anisotropies in the CMB:
the non-linear evolution of gravitational structures
(Rees \& Sciama 1968) and the mentioned SZ effect.
The latter is expected to give the largest contribution,
and we shall center our analysis on it.
Since clusters are known to posses ionized gas at temperatures
about $10^8$K they are expected to give the largest contribution
to the SZ effect. Cocoons of radio galaxies (Yamada,
Sugiyama \& Silk 1999) or other foreground sources
subtend a very small angle and, due to beam dilution,
have a very small contribution at
the scales probed by the Tenerife experiment.
Therefore, we shall look for
correlations between the Tenerife CMB data and template
maps generated from cluster catalogues; in particular,
we use the ACO (Abell, Corwin \& Olowin 1989),
and ROSAT PSPC catalogues (Vikhlinin et al. 1998).
To test the hypothesis that the CMB signal is generated
by hot diffuse gas distributed on supercluster scales
we also include a supercluster catalogue
(Einasto et al. 1994) in our analysis. 
Finally, we also constructed template
maps from the HEAO 1 A-1 map of X-ray sources (Kowalski et al. 1984)
that should trace the distribution of hot electrons.
In Sec. 2 we describe the statistical methods used 
for comparing data on temperature anisotropies with template maps
constructed from different surveys. In Sec. 3 we 
describe the catalogues used and how the template maps
were elaborated. Finally, in Sec. 4 we present
and discuss our main results.

\section{Statistical Method.}

At any given frequency, the CMB anisotropy map 
can be considered a superposition of contributions 
of cosmological origin, $T_{CMB}$, 
astrophysical origin, $\alpha M$, and instrument noise, $N$:
$T=T_{CMB}+\alpha M+N$.
$\alpha$ is a conversion factor -to be determined- that gives the 
amplitude of the contribution of foreground sources to the CMB 
temperature anisotropies. 
As mentioned in the introduction, the spatial distribution
of clusters should trace that of the electrons.
From cluster catalogues
we shall construct template maps of the temperature
anisotropies induced on the CMB spectrum by the hot gas traced by
clusters.  We shall name template map the term $\alpha M$.
Let us remark that we do not have a map of the distribution
of gas on the nearby Universe. Our hypothesis that
clusters trace the gas distribution means that we expect the 
autocorrelation function of the template map to be rather similar
to that of the hot gas even though we ignore the exact 
gas distribution. Therefore, and unlike  
de Oliveira-Costa et al. (1999), our analysis
shall be based on comparing correlation functions 
and not the maps themselves.

Assuming that the contribution of foreground sources 
is uncorrelated with the cosmological signal and noise in the
temperature anisotropy map, then the cross
correlation of the CMB and the template maps $C_{TM}(\theta)$
is related with the template map autocorrelation
function $C_{MM}(\theta)$ as:  $C_{TM}(\theta) = 
\alpha C_{MM}(\theta)$. A best-fit value of $\alpha$
is obtained by minimizing (Banday et al. 1996)
\begin{equation}
\chi^2 = \sum_{ij} 
[C_{TM}(\theta_i)-\alpha C_{MM}(\theta_i)] M^{-1}_{ij}
[C_{TM}(\theta_j)-\alpha C_{MM}(\theta_j)].
\end{equation}
In this expression $M_{ij}$ is the covariance matrix of
the cross-correlation functions (Ganga et al. 1993)
defined as follows: $M_{ij}=<[C(\theta_i)- <C(\theta_i)>]
[C(\theta_j)- <C(\theta_j)>]>$, with $\theta_i,\theta_j$ two
arbitrary angular separations in the sky. $C(\theta)$ is the
cross correlation of the template map and one single realization
of the observed sky. Realizations of the sky were performed
in two different ways: (a) at each measured temperature 
we add a random realization of a gaussian distributed noise
with zero mean and the variance at that point.
(b) We performed Monte Carlo CMB simulations of the Tenerife 
data drawn from a gaussian distribution
with variance $C_l=6C_2/l(l+1)$ at each multipole,
normalized to $Q_{rms-PS}=20\mu$K. 
We assumed a Harrison-Zel'dovich power spectrum
for the primordial fluctuations since, together with the previous
normalization, is a good approximation
at the scales probed by the Tenerife experiment
(Guti\'errez et al. 2000). To each point in the CMB map we add a 
realization of the noise as in (a).  In both cases,
the average $<..>$ was obtained from a thousand realizations.

The minimum-variance estimate is:
\begin{equation}
\hat\alpha = \sum_{ij}{C_{MM}(\theta_i)M^{-1}_{ij}C_{TM}(\theta_j)
	\over  \sum_{ij} C_{MM}(\theta_i)M^{-1}_{ij}C_{MM}(\theta_j)}
\label{alpha}
\end{equation}
with formal error
\begin{equation}
\sigma_{\hat\alpha}= (\sum_{ij}C_{MM}(\theta_i)M^{-1}_{ij}
			C_{MM}(\theta_j))^{-1/2} .
\label{salpha}
\end{equation}

The approach (a) described above
does not include sample variance.
We estimated the associated error bar 
by performing  a thousand Monte Carlo
realizations of the CMB sky and finding $\alpha$ from the
correlation with the template maps.
As expected, the average value of $\alpha$ was zero.
The dispersion around this mean, $\sigma_s$, is a measure
of both cosmic variance and the 
variance coming from random alignments.
On the other hand, the approach (b) includes all
contributions to the variance in the estimate of $\hat\alpha$.

\section{Data and Template Maps.}

The results of the Tenerife CMB experiments are presented
in Guti\'errez et al. (2000). The observations
were performed in two frequencies: 10 and 15 GHz
covering 5000 and 6500 square degrees, respectively.
The experiments are sensitive to multipoles 
$l= 10 - 30$ which corresponds to the
Sachs-Wolfe plateau of the radiation power spectrum.
The experiment measures strips in right ascension
separated by $2.5^o$ in declination.
The 15GHz map is made of 8 strips that
spans a region on the sky from 8h to 18h
in R.A.  and from $27.5^o$ to $45^o$ in Dec.
The 10GHz is slightly smaller  with only 5
strips running from $32.5^o$ up to $42.5^o$.
The map is in the North Galactic hemisphere
and  has a galactic latitude $b\ge 20^o$.
The experiment uses a double-differencing technique
to measure, with a $5^o$ FWHM beam, points separated 
$8.1^o$ in R.A. 
For 15GHz, the band power of the CMB signal 
is $\Delta T_l  = 30^{+15}_{-11}\mu$K, including a possible
contaminating effect due to the diffuse Galactic component.
The r.m.s. temperature anisotropy at $5^o$ is
$\sigma_{TEN, 10GHz} = 43\mu$K. At 15GHz, $\sigma_{TEN, 15GHz} = 32\mu$K.
The sensitivity at 10 and 15 GHz was $\sim 31\mu$K and $\sim 12\mu$K, 
respectively, in a beam-size region.

As explained in the previous section, we
assume that clusters trace
the spatial distribution of the hot gas. 
Cluster surveys select members according
to a given criteria. Therefore, different catalogues
have different selection biases.
For each catalogue we shall elaborate a template
map to compare with the Tenerife CMB data.
Let us briefly describe the ones that will be used.
The ACO all-sky catalogue contains 4073 rich clusters
of galaxies, each having at least 30 members within 
magnitude range $m_3$ to $m_3+2$ ($m_3$ is the
magnitude of the third brightest cluster member) and each with
redshift less than 0.2.  The HEAO 1 A-1 catalogue is essentially
a catalogue of ACO clusters with X-ray emission in the energy range
$0.5-20$keV.  For several nearby clusters the SZ effect 
has been measured (Birkinshaw 1998). Therefore, this catalogue
traces the extragalactic objects known to be sources of SZ. 
In this respect, it will be interesting to compare the 
results obtained by cross-correlating
each of these templates with the Tenerife map.
By including clusters that do not contribute significantly
to the SZ effect, we could have diluted the signature of the hot
gas in the cross-correlation between the ACO template and Tenerife.

ROSAT is a catalogue of X-ray selected objects. It includes
from poor groups till rich clusters of galaxies. These  clusters
were serendipitously detected in the ROSAT PSPC high Galactic
latitude pointed observations ($b\ge 30^o$). The satellite 
covers a large energy range ($0.1-2$keV) in the soft X-ray band.
The cluster redshifts range from $z = 0.015$
to $z > 0.5$ in the area of the sky covered by Tenerife.
The HEAO 1 A-1 and ROSAT catalogues are less sensitive
than optical catalogues to projection effects and
could detect "failed clusters" were galaxy formation
was suppressed. 

The Tenerife experiment operates on the Rayleigh-Jeans
regime and the effect of the hot electrons is to produce
a decrement on the radiation temperature.
Therefore, if the experiment has detected any contribution
from hot gas, cold spots in the data and the template maps
should be correlated.
The anisotropy depends linearly on the central electron temperature,
cluster core radius and electron density: 
${\delta T/T_o}\propto - r_cT_en_e$ (Zel'dovich \& Sunyaev 1969).
The exact relation depends on the cluster density
profile but this is of no significance since clusters 
are unresolved by the antenna.  The parameters $r_c, n_e$ 
and $T_e$ scale with the cluster mass (Bower 1997).
To elaborate a template map we assume that not
only clusters trace the gas distribution but also that
the cluster richness is a measure of the cluster mass, and, 
consequently, of its size, gas content and electron temperature.
We only had information on the richness  of ACO and HEAO 1 A-1 clusters.
For them, we constructed a template map by
assigning a number to each pixel:
zero if there was no cluster, and a contribution
proportional to the richness
$\propto (Richness)^n$ if there was a cluster.
Since we did not know how to scale the cluster size,
gas density and electron temperature 
with richness class we tried exponentiating the
richness class to three different powers:
$n=0,1,2$. Finally, this pixel map was
convolved with the Tenerife window function.
To avoid boundary effects, we included objects within
$15^o$ of the region probed by Tenerife. 
Like in Bennett et al. (1993) we found no significative
effect: while for the ACO clusters larger $n$ led to 
diminishing the cross-correlation and consequently
the value of $\alpha$, for the HEAO 1 A-1 catalogue, the
opposite effect was observed.
However, in all cases the effect was minute and 
well within the error bars. 
We shall quote our results for $n=0$, when all clusters
contribute equally to the SZ effect. 
The template map of the ROSAT catalogue was constructed
in a similar manner but without scaling with richness class.

Finally, we also included in our analysis
the supercluster catalogue of Einasto et al. (1994).
This catalogue was elaborated from the distribution 
of rich clusters of galaxies
up to redshift $z=0.1$, extracted from the ACO catalogue
described above.  For each supercluster, the number
of cluster members, center position, average distance $D$, extent
in supergalactic coordinates and length $L$
in Mpc are given. In our analysis, twenty two superclusters with
typical angular sizes between 5 and $10^o$ were included. 
Contrary to clusters, superclusters can not be considered 
point-like. Therefore, different hypothesis
about the gas distribution could lead to different
results.  As a first approximation, 
we took the gas distributed homogeneously on a 
sphere of size $2\tan^{-1}(L/2D)$. 
This template is very convenient in order to check Hogan (1992)
hypothesis about the local origin of temperature
anisotropies. We called this template "superclusters
with homogeneous gas distribution".
We checked that the correlation level did not depend
on the gas distribution by chosing a
model with a density profile:
\begin{equation}
n(r) = {n_e\over 1+(r/r_c)^2}
\end{equation}
where $r_c\sim L/10$ is a fiducial radius. 
We called this template "superclusters with concentrated gas".
Finally, both templates were
convolved with the Tenerife beam pattern before performing
the correlation analysis.

\section{Numerical Results and Discussion.}

Table 1 summarizes the results of our analysis.
After substracting the mean and normalizing the templates
to unit variance, the autocorrelations and cross correlations 
were computed given equal weight to each pixel. 
We tried different angular bins, from 1$^o$ to 5$^o$, and computed
the correlation function out to $20^o$ and $30^o$. No
significant differences were found.  
In Table 1, the results are quoted for correlation
functions in bins of $3^o$ out to $21^o$.

The SZ effect will generate approximately equal and negative
contributions at 10 and 15GHz. If the signal detected
is real one should expect 
equal and positive values of $\alpha$ at the two frequencies.
In Table 1 we give $\hat\alpha$,
$\sigma_{\hat\alpha}$ as given by eqs.~(\ref{alpha}) and
(\ref{salpha}), and $\chi^2$ per degree of freedom (dof)
in the two approaches described in Sec. 2.
For simplicity, we termed (a) ``without sample variance'' 
and (b) ``with sample variance''.
In the case (a), we also give the error associated
with sampling variance $\sigma_s$, which should be added
in quadrature with $\sigma_{\hat\alpha}$. 
As the maps were normalized to unit variance, $\alpha\sigma_{TEN}$
gives the SZ component of the CMB map in thermodynamic units.
For each template we calculate the cross-correlation
with the 15GHz and 10GHz maps. Since the latter
covers a smaller fraction of the sky, we also correlate a 
reduced 15GHz map (denoted by 15c in the table) 
cut to the size of the 10GHz map
to eliminate the bias introduced by the different
sky coverage. 

No significative detections (larger than $2\sigma$) were found by either 
of the two methods. We always found negative values of $\alpha$, i.e.,
cold spots in the template map correlate with hot spots in the data, contrary 
to what one would expect if there was a significant SZ contribution.
The largest signal was obtained at the 
15c ROSAT template map. Since the amplitude of the SZ 
effect does not change much at the Tenerife frequencies,
consistency would require
a fluctuation of the same order to be present at 10GHz.
Furthermore, when sample variance was not included in the covariance
matrix, the best-fit was never a good fit. Only
when it was included $\chi^2/dof$ became of order unity. 
The low quality of the fit can be understood by looking at Figure 1, where
we plot the autocorrelation of the template maps (Fig 1a)
and their cross correlation with the data on 15GHz (Fig 1b).
The dashed line corresponds to the autocorrelation and
cross correlation of the ACO catalogue, the long dashed line
to the HEAO 1 A-1 catalogue, the dot-dashed line to 
the ROSAT catalogue. Thick and thin solid lines correspond
to superclusters with uniform and concentrated gas, respectively.
While the autocorrelation functions are
rather similar in all cases, the cross-correlation differ
substantially in shape.  When the sample variance
is not included in the covariance matrix, $\sigma_{\hat\alpha}$
is small and the difference in shape implies a large
$\chi^2/dof$. Only when sample variances are included, 
the error bars are much larger and the fit to the data improves.
To conclude, we can only set upper limits on the value of
$\alpha$. Taking the results on 15GHz, we limit $\alpha\le 0.24$
at the 99\% confidence level.

For an experiment with such a beam width like Tenerife, 
one could not expect to
find a large correlation between data and templates.
For example, some clusters in the ACO catalogue have
been found to produce temperature fluctuations
of the order $100\mu$K (Birkinshaw 1999). But they subtend an
angular scale of few arcminutes and as a result the SZ signal is 
diluted by the large solid angle covered by the Tenerife beam.
Still, the Tenerife data limits the
contribution of nearby clusters and superclusters
to be smaller than $8\mu$K at 99\% confidence level.
The mean Comptonization parameter at $y = {-\Delta T\over 2T_o}
\le -1.5\times 10^{-6}$ at the same level of
confidence. Our results are slightly more
restrictive than those previously found by Banday et al. (1996). 
Let us remark that $y$ obtained above only limits 
the contribution due to nearby superclusters, while
the COBE result of  Mather et al. (1994) applies to the contribution of 
all structures located between the last scattering surface and the observer.
To conclude, this study, like previous ones based on the COBE/DMR
data, indicate that most of the signal measured by Tenerife is 
not of extragalactic origin but cosmological.

\acknowledgments

We thank R. Rebolo for many useful discussions and comments.
F.A.B acknowledge the financial support of the University
of La Laguna - Banco de Santander. F.A.B. and C.H.M. acknowledge
the hospitality of the I.A.C. where most of this work was 
carried out.

\newpage
\begin{deluxetable}{lccccccc}
\tablewidth{33pc}
\tablecaption{Cross-correlation results.}
\tablecolumns{8}
\tablehead{  
\colhead{}    &  \multicolumn{4}{c}{without sample variance} &  
\multicolumn{3}{c}{with sample variance} \\  
\cline{2-5} \cline{6-8} \\  
        \colhead{$\nu$/GHz}
&       \colhead{$\hat\alpha$} 
&       \colhead{$\sigma_{\hat\alpha}$}
&       \colhead{$\sigma_s$}
&       \colhead{$\chi^2_{min}/dof$}
&       \colhead{$\hat\alpha$} 
&       \colhead{$\sigma_{\hat\alpha}$}
&       \colhead{$\chi^2_{min}/dof$}
        }       
\startdata
\cutinhead{ACO Clusters}
15&     -0.05&    0.017&  0.08&  17&  -0.03&  0.07&  1.1\nl
15c&    -0.019&    0.02&  0.09&   4&  -0.03&  0.1 &  0.3\nl
10&     -0.06&     0.03&  0.09&   2&  -0.05&  0.07&  0.4\nl
\cutinhead{SUPERCLUSTERS (concentrated gas)} 
15&     -0.12&   0.015&   0.09&  13& -0.05&  0.06&  1.6\nl 
15c&    -0.09&    0.02&   0.09&  11& -0.05&  0.07&  1.4\nl
10&     -0.10&    0.02&   0.09&  11& -0.1 &  0.06&  1.8\nl
\cutinhead{SUPERCLUSTERS (homogeneously distributed gas)} 
15&     -0.11&    0.017&   0.09& 12&  -0.05&  0.06&  1.4\nl
15c&    -0.10&     0.02&    0.1& 11&  -0.06&  0.07&  1.3\nl
10&     -0.09&     0.02&    0.1& 7&   -0.08&  0.07&  1.4\nl
\cutinhead{HEAO 1 A-1 X-ray sources}
15&     -0.018&   0.016&  0.07& 17& -0.02&  0.06&  1.7\nl
15c&    -0.06&    0.02&   0.08& 10& -0.08&  0.07&  1.16\nl
10&     -0.02&    0.02&   0.08&  7& -0.014& 0.06&  0.9\nl
\cutinhead{ROSAT clusters.}
15&     -0.12&    0.018&   0.09& 11&	-0.15&  0.06&  1.3\nl
15c&    -0.17&    0.02&    0.1&  20&	-0.19&  0.07&  2.1\nl 
10&     -0.04&    0.02&    0.1&   2& 	-0.04&  0.06&  0.3\nl
\enddata
\end{deluxetable}

\newpage
\begin{figure}[ht]
\plotone{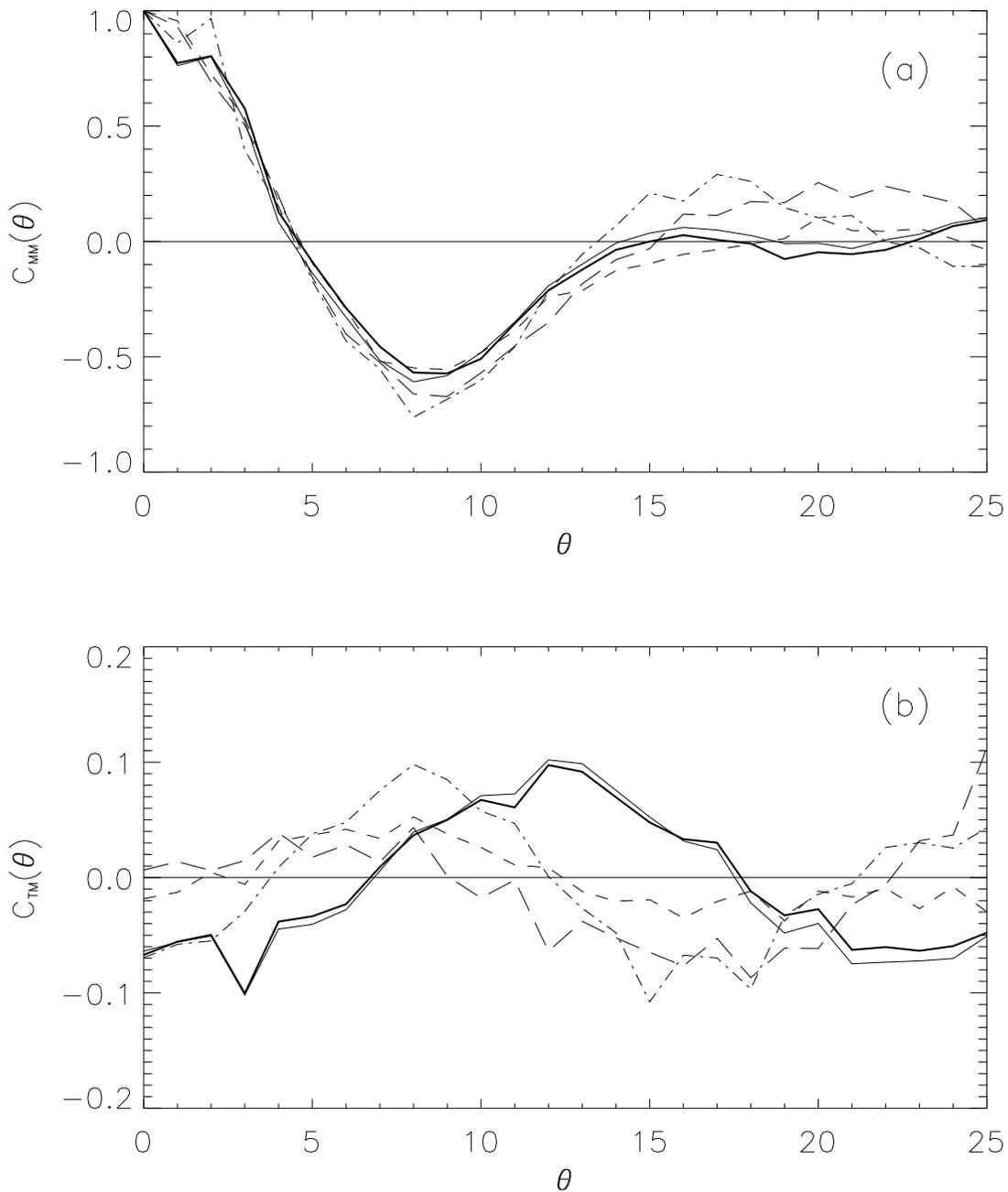}
\caption{(a) Autocorrelation of template maps normalized
to unit variance.
(b) Cross-correlation of template maps with the
15GHz CMB data. The x-axis is measured in degrees. In both plots, 
dashed lines correspond the ACO catalogue, the long dashed lines
to the HEAO 1 A-1 catalogue, the dot-dashed lines to 
the ROSAT catalogue. Thick and thin solid lines correspond
to superclusters with uniform and concentrated gas distributions,
respectively. Templates and CMB data were normalized to zero mean and unit
variance.
}
\label{fig1}
\end{figure}

\end{document}